# Lateral Confinement of Electrons in Vicinal N-polar AlGaN/GaN Heterostructure

**Digbijoy N. Nath**[1a)], **Stacia Keller**[2], **Eric Hsieh**[2], **Steven P. DenBaars**[3], **Umesh K. Mishra**[2] **and Siddharth Rajan**[1]

[1] Department of Electrical and Computer Engineering, The Ohio State University, Columbus OH 43210 USA
[2]Electrical and Computer Engineering Department, [3]Materials Engineering Department
University of California, Santa Barbara, Santa Barbara CA 93106 USA

**Keywords**: Vicinal, N-polar, atomic steps, Quasi 1-D channel

## Abstract

We studied orientation dependent transport in vicinal N-polar AlGaN/GaN heterostructures. We observed significant anisotropy in the current carrying charge parallel and perpendicular to the miscut direction. A quantitative estimate of the charge anisotropy was made based on gated TLM and Hall measurements. The formation of electro-statically confined one-dimensional channels is hypothesized to explain charge anisotropy. A mathematical model was used to verify that polarization charges distributed on miscut structure can create lateral one-dimensional confinement in vicinal substrates. This polarization-engineered electrostatic confinement observed is promising for new research on low-dimensional physics and devices besides providing a template for manufacturable one-dimensional devices.

---

[a)] Author to whom correspondence should be addressed. Electronic mail: nathd@ece.osu.edu
Tel: +1-614-688-8458

Recent work has shown that it is possible to achieve good microwave performance using high electron mobility transistors (HEMTs) on $(000\bar{1})$ oriented (N-polar) GaN.[1,2,3] Initial results were achieved by molecular beam epitaxy (MBE) due to challenges related to N-polar growth by metal-organic chemical vapor deposition (MOCVD). It had been however reported that MOCVD growth on vicinal GaAs led to naturally formed multi-atomic steps[4]; growth of fractional layer super-lattice too had been shown to lead to pronounced step-ordering[5] on vicinal <001> GaAs. Surface instability issues related to MOCVD growth of N-polar GaN were thus resolved by growth on vicinal sapphire substrates[6,7] where surface steps induced by the miscut were found to improve surface mobility and mitigate three-dimensional growth and formation of hexagonal pyramids on the surface. This has led to excellent structural and surface quality for MOCVD grown N-polar GaN films with good DC and RF transistor performance[8].

Previous results of vicinal *Ga-polar* AlGaN/GaN heterostructures showed anisotropy in mobility of electrons parallel and perpendicular to the steps[9,10]. The difference in mobility in the two directions was attributed to the anisotropy in roughness, with an enhancement in the mobility along the step direction. With the development of N-polar GaN by MOCVD, mobility and device characteristics of MOCVD-grown GaN/AlGaN/GaN structures were investigated.[7] Brown *et al*[11] studied conduction in the parallel direction and explained the vertical field dependence of the mobility. However, there is no previous detailed investigation of anisotropy in the transport properties of N-polar vicinal structures. Although observations of anisotropy in mobility[12] and in transistor current (Id-Vd)[13] for vicinal GaAs-based heterostructures were attributed to probable step-related lateral potential modulation,[12] however there has been no mathematical modeling of this potential landscape to our knowledge. Besides, anisotropy in the 2DEG charge in the form of anisotropy in the pinch-off of transistor channels is not reported. In this work, we used transistor and gated TLM measurements to show that there is anisotropy in the conductive charge density of the electron gas parallel and perpendicular to the miscut-induced steps. We provide a simple theoretical model to qualitatively explain our experimental results.

The epitaxial structure used in this study was grown by metal organic chemical vapor deposition (MOCVD) on a sapphire substrate mis-oriented 3° towards the a-plane. Details of semi-insulating and smooth buffer growth can be found in previous work [6]. Closer analysis of the surface using AFM revealed step bunching, and the step height was typically 2-4 monolayers (5 – 10 Å). The active layer used in the experiment consisted of an $Al_{0.33}Ga_{0.67}N$ barrier, followed by a 30 nm GaN channel layer. The layers were capped by 5 nm SiN insulator grown by MOCVD after growth. The SiN layer was added to improve surface stability and to simplify lithography and device fabrication. Device fabrication was carried out on the epitaxial layer described above. Several transistors, Schottky diode, gated TLM, and Hall patterns were fabricated on the epitaxial layers. Standard i-line stepper lithography was used to define patterns. The ohmic contact consisted of a Ti/Al/Ni/Au (20/100/10/50 nm) stack deposited by e-beam evaporation. This was followed by $Cl_2$-based reactive ion etching for mesa isolation. Following gate lithography, e-beam deposition of Ni/Au/Ni (30 nm/300 nm/30 nm) gate metal stack was done. Using the stepper, alternate dies (each with several patterns) were oriented parallel and perpendicular to the steps. To study the dependence of orientation on transport, adjacent dies were compared so that the effect of growth or process variations across the sample was minimal.

A Keithley 4155 Semiconductor Parameter Analyzer was used, and a 5-probe tip measurement technique was employed with four probes for the resistance measurement, and a fifth for applying the gate bias. The resistance of TLM patterns with gate lengths varying from 2 to 20 μm and identical access regions was measured. The slope of the conductance with respect to the gate length was used to calculate the channel conductivity at each gate bias. The current was lowered until there was no dependence of conductivity on the current level, implying that we could ignore high-field effects and pinch-off in the channel.

In figure 1, we show the transconductance ($g_m$–$V_{GS}$) for two transistors (parallel, A and perpendicular, B to the step direction) with a gate length of 10 μm. The pinch-off voltage of transistor **A** is higher in magnitude than that for transistor **B** by approximately 1 V. We found that the difference in pinch-off voltage between perpendicular and parallel devices was maintained for drain biases up to 40 V, gate lengths down to 0.7 μm, and temperature up to 300°C which indicated that the charge density available for transport along the parallel direction is higher than that available for transport perpendicular to the steps. This is confirmed by Hall Effect measurements performed by passing current parallel and perpendicular to the direction of the steps which yielded a Hall charge density of $1.37 \times 10^{13}$ cm$^{-2}$ and $1.2 \times 10^{13}$ cm$^{-2}$ for current flow in directions parallel and perpendicular to the steps respectively. The inset to Figure 1 shows the capacitance-voltage (C-V) profiling of the charge in the two directions, indicating clearly the anisotropy in charge as verified by Hall measurements.

The mobility $\mu(V_G)$ for an applied gate bias $V_G$ was calculated using the equation

$$\mu(V_G) = \frac{1}{qR_{sh}(V_G)n_s(V_G)}$$

where $R_{sh}(V_G)$ is the sheet resistance, and $n_s(V_G)$ is the charge deduced from gated TLM and C-V measurements, respectively. The mobility thus obtained is found to be anisotropic (Figure 2). However, since the charge density is anisotropic in two directions, mobility is 'normalized' using an approximation to estimate effective charge at each gate bias. We express the charge as

$$n_{S,X}(V_G) = C_X(0)(V_G - V_{P,X})$$

where 'X' represents the direction of transport (parallel or perpendicular), $C_X(0)$ is the zero-bias capacitance of the fat-FET pattern measured in the direction 'X', and $V_{P,X}$ is the pinch-off voltage in the direction 'X'. The mobilities for the parallel and perpendicular directions calculated by this method are found to be similar after considering the anisotropy in available charge (inset to Figure 2).

We developed a simple electrostatic model (Figure 3) to qualitatively explain the anisotropic charge transport measured in this work. The model approximates the vicinal steps by uniform terraces of width d = 20 nm and height h = 1 nm (giving a miscut angle of approximately 3°). An infinite number of steps is assumed so that the potential profile corresponding to each step is the same. It provides calculation of an energy band landscape. Only the polarization charge has been considered; adding electrons in the form of 2DEG would change the profile.

The sheet charge at each vicinal step is assumed to be composed of line charges of infinitesimal thickness dx. In the x-z plane associated with each step where z-axis is the growth direction, at any point $(x_0,z_0)$, the electric field E due to $\sigma_P$ (the net positive polarization charge at GaN/AlGaN interface) at the $N^{th}$ step (from the step considered) is composed of x and z components given by

$$E_x(N) = \int_{x=0}^{x=d} \frac{\sigma_P \cos\theta \, dx}{2\pi\varepsilon_r\varepsilon_0 r} \qquad (1)$$

$$E_z(N) = \int_{x=0}^{x=d} \frac{\sigma_P \sin\theta \, dx}{2\pi\varepsilon_r\varepsilon_0 r} \qquad (2)$$

where 'r' is the distance of the line charge considered from the point $(x_0,z_0)$. The net electric field in x and z directions are therefore obtained by summing $E_x(N)$ and $E_z(N)$ for a large number of steps N. The contributions of the different charges at the different interfaces can be incorporated similarly. The energy band landscape associated with each step in the x-z plane can be plotted by evaluating potential $V(x_0,z_0)$ as

$$V(x_0,z_0) = \int_{(x=0,z=0)}^{(x=x_0,z=0)} E_x \, dx + \int_{(x_0,z=0)}^{(x_0,z=z_0)} E_z \, dx \qquad (3)$$

The potential profile for a single step calculated using this model shows a lateral confinement potential of ~ 250 meV (inset to Figure 3) in the x-direction at a vertical distance of $z_0 = 2$ nm from the GaN/AlGaN interface due to the step polarization charge for a GaN thickness of 10 nm and AlGaN thickness of 20 nm. The conduction band for the overall step structure assumes a saw-tooth profile (inset to Figure 3). As the channel density is reduced by applying a negative gate bias, the Fermi level of the electrons drops below the conduction band and no conduction can occur across the steps, while the channel is still conductive parallel to the steps. From our analysis, the magnitude of the confining potential was proportional to the step height and the polarization sheet charge density. The step length did not change the depth of the confining potential.

The lateral polarization engineering described here can be used to create quasi one-dimensional potential profiles without lateral patterning or epitaxy. It provides a method to achieve lateral quasi one-dimensional channels with electrostatic confinement, rather than quantum mechanical heterostructure-based confinement. Furthermore, the polarization charge density (alloy composition) and vicinal miscut angles can be varied to achieve higher or lower confinement to engineer the spatial location, density, and other properties of the channel.

In conclusion, we studied orientation dependent transport in vicinal N-polar AlGaN/GaN

heterostructures. We observed significant anisotropy in the current carrying charge parallel and perpendicular to the miscut direction. A quantitative estimate of the anisotropy in charge was made using a combination of Hall and capacitance-voltage measurements. The formation of quasi one-dimensional electrically confined channels is hypothesized to explain the charge anisotropy. A mathematical model was used to verify that polarization charges can indeed create lateral quasi one-dimensional confinement in vicinal substrates. The self-patterned one-dimensional carrier channels using this method could be used not only to study interesting and new physics of low-dimensional systems, but also as a template for manufacturable one-dimensional devices.

We thank Prof. Debdeep Jena for useful discussions on the theoretical analysis. Funding from ONR (Dr. Paul Maki) and OSU IMR is gratefully acknowledged.

**Figure Captions**

**Figure 1**: (color online) The anisotropy in transconductance and thus in pinch-off for current flowing in directions parallel (A) and perpendicular (B) to steps. Inset: C-V profiling of the charge

**Figure 2**: Anisotropy in mobility as measured using Hall pads and gated TLMs. Inset: Mobility after taking into account the anisotropic charge density.

**Figure 3**: (color online) Electrostatic model involving polarization charge at vicinal GaN/AlGaN interface to explain the anisotropy of charge distribution in this study. Inset: Saw-tooth energy band profile for lateral direction.

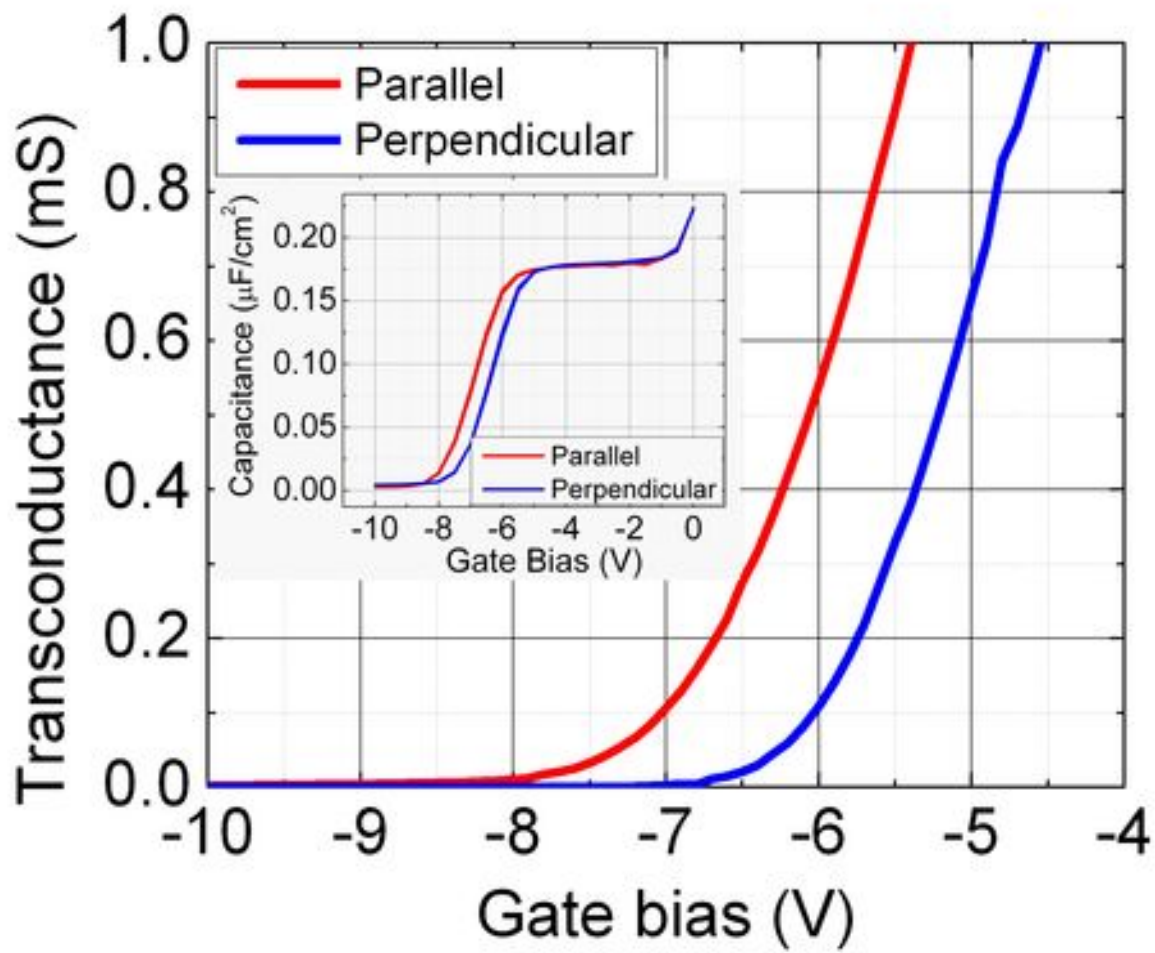

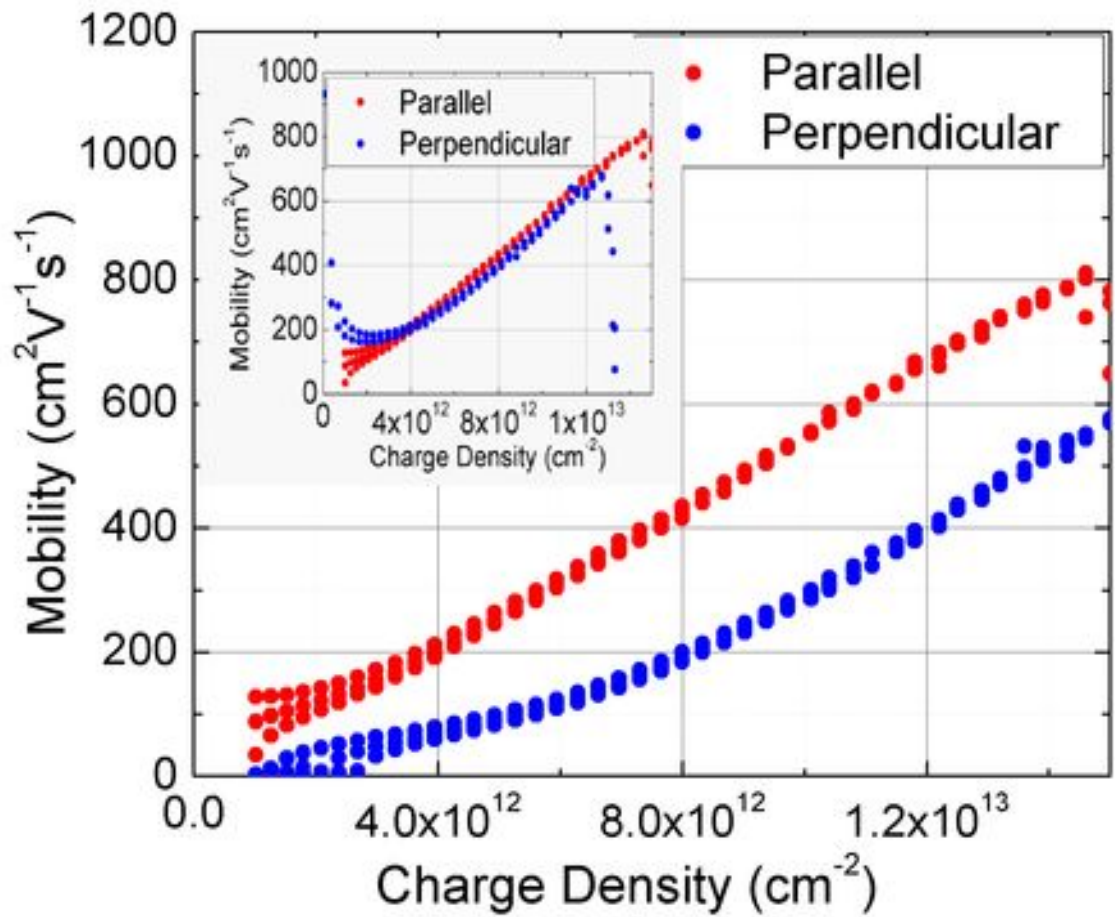

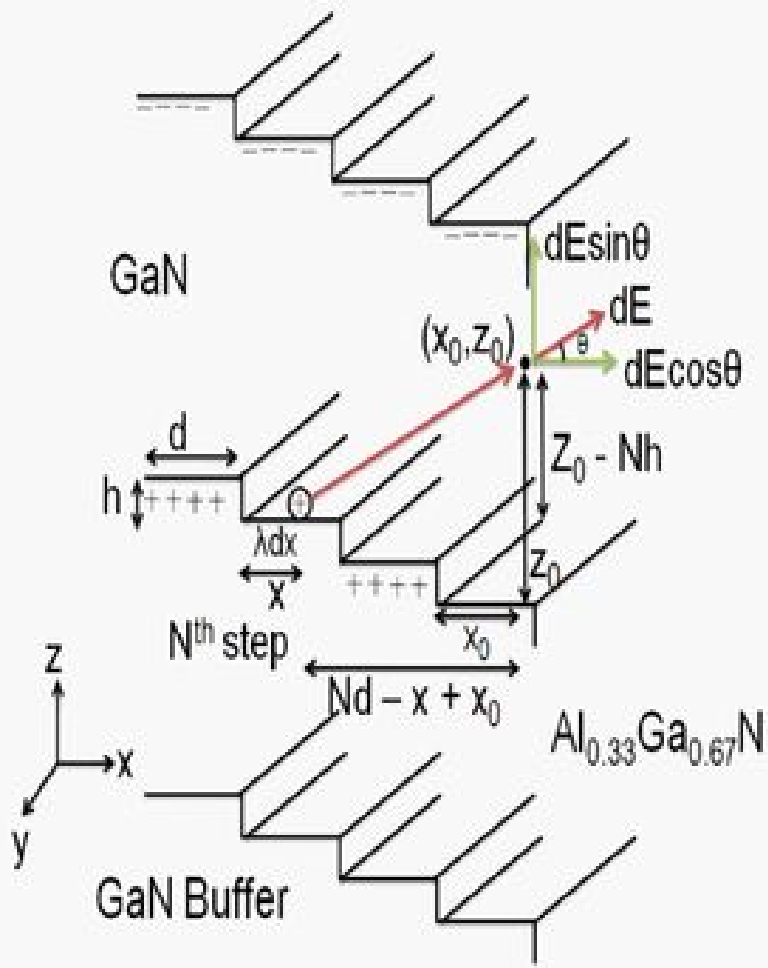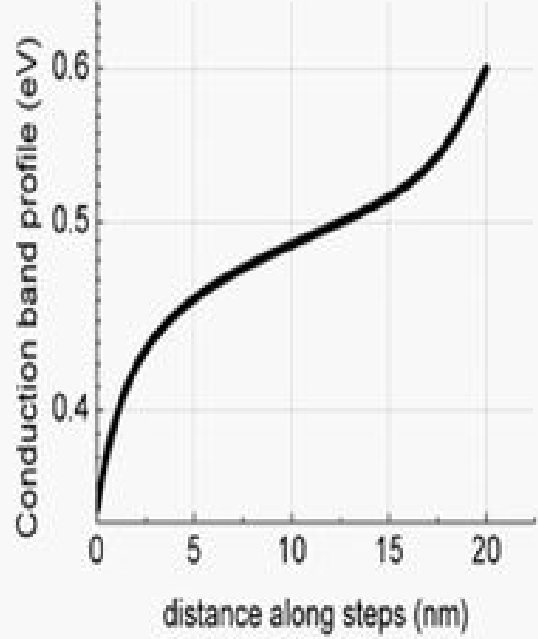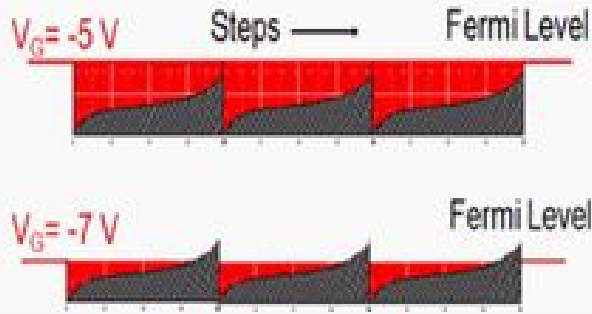